\documentclass[galaxies,conferencereport,accept,moreauthors,submit,10pt,a4paper]{mdpi}

\usepackage{amsmath}

\usepackage{booktabs}
\usepackage{multirow}
\usepackage{soul}
\usepackage{microtype}
\firstpage{1}
\articlenumber{42}
\doinum{10.3390/------}
\pubvolume{5}
\pubyear{2017}
\copyrightyear{2017}
\externaleditor{Academic Editors: Duncan A. Forbes; Ericson D. L\'opez}
\history{{Received: 23 June 2017; Accepted: 15 August 2017; Published: 18 August 2017}}

\pdfoutput=1

\Title{The Baryonic Halos of Isolated Elliptical Galaxies}

\Author{{Ricardo Salinas} $^{1,}$*,  
Adebusola {Alabi} $^{2,3}$, Nicklas {Hammar} $^{1,4}$,  
Tom Richtler $^{5}$, Richard R. Lane $^{6}$ and Mischa Schirmer $^{1}$}

\AuthorNames{Ricardo Salinas, Adebusola Alabi, Nicklas Hammar, Tom Richtler, Richard R. Lane, Mischa Schirmer}

\address{%
$^{1}$ \quad Gemini Observatory, Colina el Pino s/n, Casilla 603, {La Serena 1700000},  
Chile; nhammar@live.ca (N.H.); mschirmer@gemini.edu (M.S.)\\
$^{2}$ \quad Centre for Astrophysics \& Supercomputing, Swinburne University, Hawthorn VIC 3122, Australia; aalabi@swin.edu.au\\
$^{3}$ \quad University of California Observatories, Santa Cruz, CA 95064, USA\\
$^{4}$ \quad Department of Physics and Astronomy, University of Victoria, Victoria, BC V8W 3P2, Canada\\
$^{5}$ \quad Departamento de Astronom\'ia, Universidad de Concepci\'on, Concepci\'on 3349001, Chile; tom@astroudec.cl\\
$^{6}$ \quad Instituto de Astrof\'isica, P. Universidad Cat\'olica de Chile, {Santiago 7820436}, Chile; rlane@astro.puc.cl}
\corres{Correspondence: rsalinas@gemini.edu}

\abstract{\textls[-15]{Without the interference of a number of events, galaxies may suffer in crowded environments} (e.g., stripping, harassment, strangulation); isolated elliptical galaxies provide a control sample for the study of galaxy formation. We present the study of a sample of isolated ellipticals using imaging from a variety of telescopes, focusing on their globular cluster systems as tracers of their stellar halos. Our main findings are: (a) GC color bimodality is common even in the most isolated systems; (b) the specific frequency of GCs is fairly constant with galaxy mass, without showing an increase towards high-mass systems like in the case of cluster ellipticals; (c) on the other hand, the red fraction of GCs follows the same inverted V shape trend with mass as seen in cluster ellipticals; and (d) the stellar halos show low S\'ersic indices which are consistent with a major merger origin.  }

\keyword{galaxies: star clusters: general; galaxies: halos; galaxies: elliptical and lenticular; cD}

\conferencetitle{On the origin (and evolution) of baryonic galaxy halos}

\begin{document}


\section{Introduction}

 As tracers of star formation, galaxy assembly, and mass distribution, globular clusters have provided important clues to our understanding of the formation and structure of early-type galaxies. However, their study has been mostly confined to galaxy clusters (e.g., \cite{peng08}), leaving the properties of the globular cluster systems (GCSs) of isolated ellipticals as a mostly uncharted territory.


      Having poorer merger histories, isolated ellipticals are particularly relevant to understanding the environmental influence on the formation of a GCS. In hierarchical-merging inspired models, the~properties of GCSs and dark halos of host galaxies in low-density environments should be very different from their high-density counterparts \cite{niemi10,tonini13}. In particular, the slope of the outer stellar halo profile will differ depending on the accretion history of each galaxy (e.g., \cite{abadi06}).

In this contribution, we present results from our ongoing work on isolated ellipticals, focusing on the properties of their GCSs and giving preliminary results on deep imaging of their stellar halos.


\section{Globular Clusters in Isolated Ellipticals}
\vspace{-8pt}
\subsection{Observations}

Targets observable from the southern hemisphere were selected from the isolated ellipticals catalogues of \cite{reda04,smith04} .

Observations were conducted using CTIO/MOSAIC-II (in the $C$ and $R$ filters,~NGC 5812, NGC~3585, and NGC 7507), Gemini-S/GMOS ($g$ and $i$ filters, NGC 2271, NGC 2865, NGC 3962, NGC 4240, and IC 4889) and VLT/VIMOS ($B$ and $R$; NGC 720, NGC 821, NGC 1162, NGC 7796, and ESO-194G021). In the present contribution we compile the results for nine of these ellipticals. Expanded details of these results have been given in \cite{caso13,lane13,richtler15,salinas15}.

Photometry of GCs was performed with DAOPHOT/ALLSTAR \cite{stetson87} after subtraction of a model of the parent galaxy in each image with IRAF/ellipse. We selected GCs  based on their point-source appearance and colors consistent with old single stellar populations.

\subsection{Results}

Color bimodality is one of the main features of GCSs, and our work shows that isolated ellipticals are not an exception. Of the nine galaxies studied, six present clear bimodality, while the unclear status of the remaining three can probably be attributed to a population of intermediate age, which, based on the predictions from stellar population models,  is expected to have optical colors between the classic red and blue globular clusters, hence blurring the bimodality signal.

\begin{figure}[H]
\centering
\includegraphics[width=0.5\textwidth]{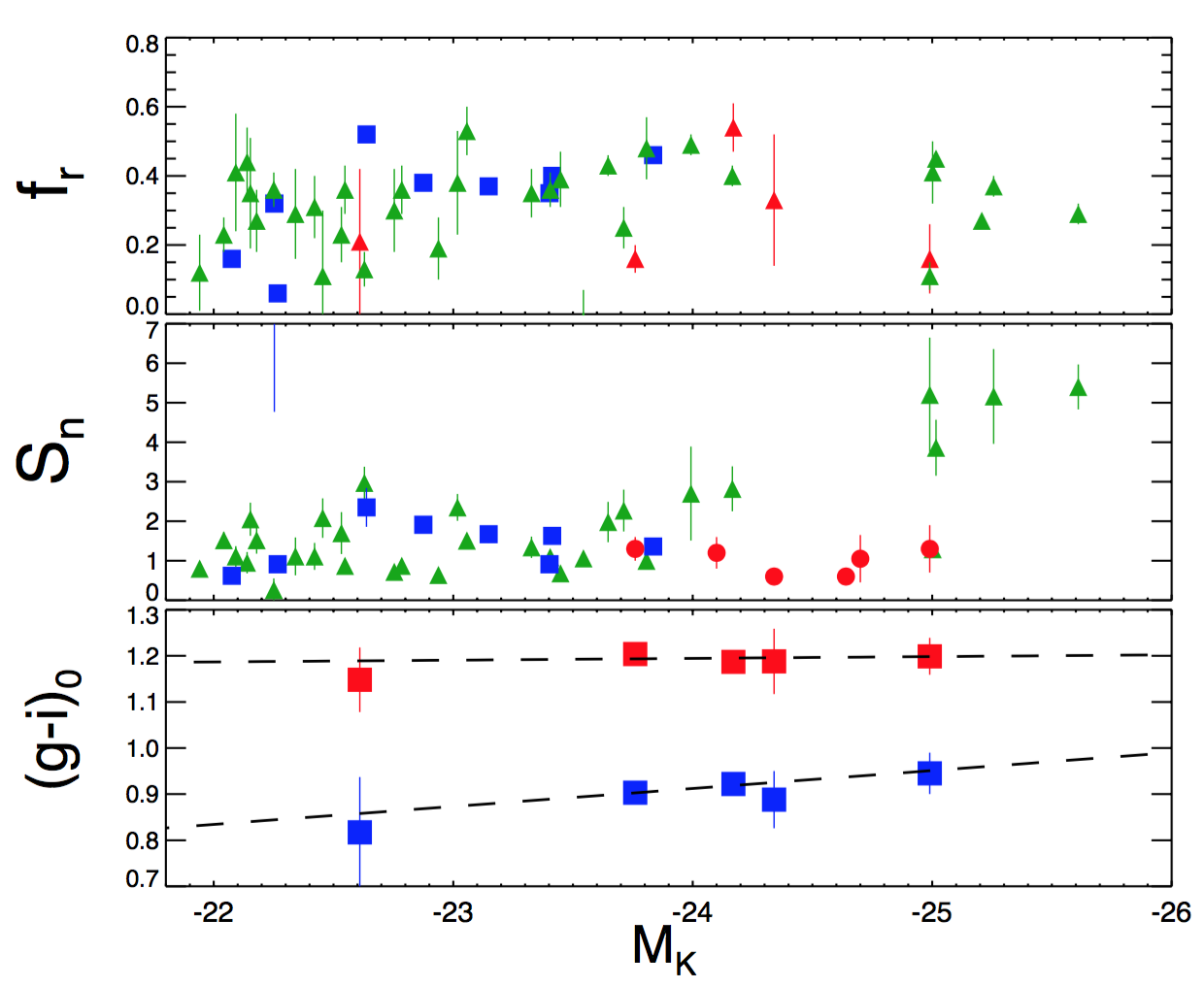}
\caption{{\textbf{{Upper panel}}}: Globular cluster (GC) red fraction as function of 2MASS $K$-band luminosity, as~a~proxy for stellar mass. Red symbols represent our sample, while blue squares are taken from the \cite{cho12} sample of 10 ``low-density'' early-type galaxies. Green triangles are the results from the ACS Virgo ClusterSurvey~\cite{peng08}; {\textbf{{Middle panel}}}: GC specific frequency. Same symbols, including the isolated ellipticals N7507 \cite{caso13}, N3585 and N5812 \cite{lane13}; {\textbf{{Lower panel}}}: position of the red and blue peaks as a function of $K$ luminosity for the five isolated ellipticals in \cite{salinas15}.}  
\label{fig:photom}
\end{figure}

A second diagnostic for the formation of a GCS is the red fraction; that is, the number of metal-rich over the total number of clusters (e.g., \cite{peng08}). The models presented in \cite{tonini13} predict that the red fraction will be a function of the accretion history; while galaxies with a rich accretion history are expected to be dominated by blue clusters (and hence have a low red fraction), the opposite is expected for galaxies with a poor accretion history.  Our measurement of the red fraction in our sample, compared to galaxies in low-density environments and in Virgo can be seen in Figure \ref{fig:photom} (top panel). The red fraction of field (blue symbols) and isolated ellipticals (red symbols), cannot be distinguished from Virgo ellipticals (green symbols), contradicting the naive expectation from the models of \cite{tonini13}.

Another useful diagnostic is the specific frequency of GCs, $S_N$, which measures the efficiency of cluster formation compared to the stellar population.~Figure \ref{fig:photom} (middle panel) shows another comparison between field, isolated, and Virgo ellipticals. The most relevant feature is the constancy of $S_N$ for field/isolated ellipticals with $S_N\sim1.5$, regardless of the host galaxy mass.~On the other hand, Virgo~ellipticals show a departure from field ellipticals starting at $M_K=-24$, which only accentuates at brighter luminosities.~Finally, the bottom panel indicates the peak color of the red and blue sub-populations as a function of luminosity. Even though the slope for the blue peak is consistent with cluster ellipticals, we do not yet have an explanation for the flat behavior of the red peak.


\section{Deep Imaging of Isolated Ellipticals}

The results from GCSs of isolated ellipticals give inconclusive evidence about the assembly of those galaxies. While bimodality and red fractions are similar to the ones in cluster ellipticals, hinting at a GCS formation process independent of environment,  $S_N$s indicate that some process must be different, at least in the bright end. Was their isolation a common feature throughout their lifetimes implying a poor merger history, or is it only a transient feature, in the sense that their isolation only arises from a rich accretion history where all the environment has been merged to form this single~galaxy?

One way to glean insight into the accretion history of galaxies is by measuring their outer stellar profile. Galaxies which have experienced a rich accretion history are expected to have an extended outer profile (formed by accreted material), while the stellar halo of truly isolated systems is expected to be sharply truncated at some radius (e.g., \cite{abadi06}). The low $S_N$ means GCSs of these galaxies will be relatively poor, and testing this prediction relies not on the outer distribution of GCs, which is necessarily deficient when based only on optical colors, but on the diffuse stellar component itself.

\subsection{Observations}

Imaging for six ellipticals in our isolated sample was conducted with two of the 1 m telescopes within the Las Cumbres Observatory network, with total on-source exposures close to 4 h per target. The Sinistro cameras give a field-of-view of $27\times27$ arcmin$^2$, which allows the construction of night-sky frames directly from the dithered science frames. Reduction of the images was conducted using the general imaging pipeline THELI \cite{schirmer13}.

\subsection{Results}

\begin{figure}[H]
\centering
\includegraphics[width=0.35\textwidth]{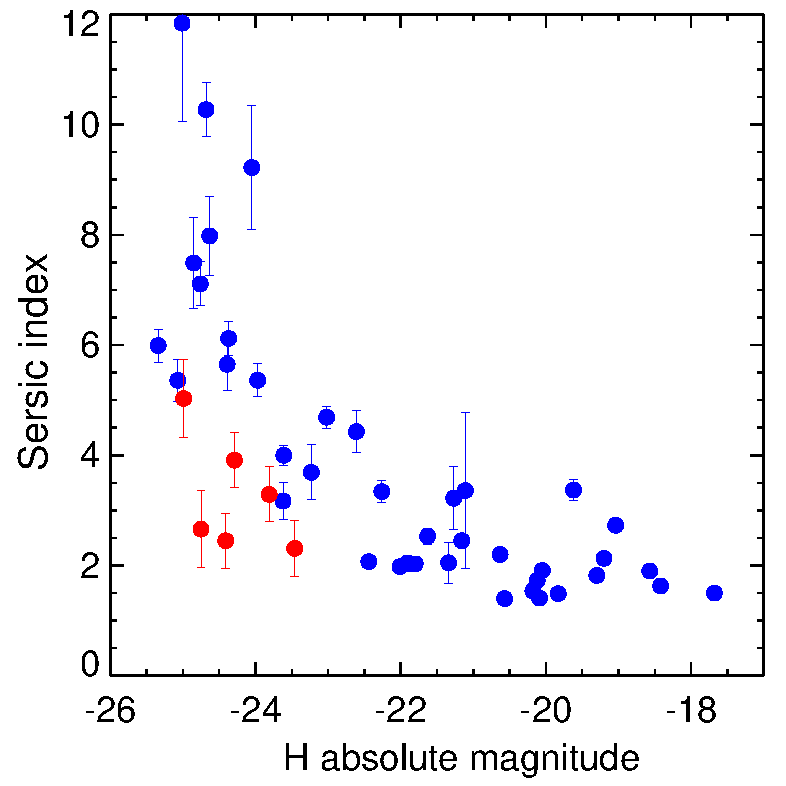}
\caption{S\'ersic  index, $n$, as function of $H$ absolute magnitude derived from 2MASS photometry. Blue~circles show Virgo ellipticals from \cite{kormendy09}, while red symbols are our Las Cumbres sample. }
\label{fig:sersic}
\end{figure}

Surface brightness profiles of the galaxies were measured with IRAF/ellipse after an iterative masking of the surrounding sources. The surface brightness profiles were then modeled with a S\'ersic profile. Figure \ref{fig:sersic} shows a comparison of the S\'ersic index of the galaxies in our sample compared to galaxies in \cite{kormendy09}. At the same $H$ absolute magnitude (used as a proxy mass), isolated ellipticals show lower S\'ersic indices, (between 2 and 5) compared to Virgo ellipticals from \cite{kormendy09} (with $n$ up to 12). This~is an indication that isolated ellipticals have experienced less mergers and are probably a product of only a major merger.

\section{Conclusions and Outlook}

\begin{itemize}[leftmargin=*,labelsep=5.8mm]
\item Optical color bimodality is a common feature in GCSs, even for the most isolated elliptical~galaxies.
\item The red fractions of GCs in isolated ellipticals are on average lower than the ones in cluster environments, at odds with the predictions of hierarchical merging (e.g., \citep{tonini13}).
 Specific frequencies are also lower compared to cluster galaxies, with a remarkably flat behavior around $S_N\sim 1.5$.
\item The blue peaks of GCSs in isolated ellipticals present a mild correlation with galaxy luminosity, although this result is driven mostly by the faintest galaxy in the sample (NGC 4240).
\item Deep imaging of six isolated ellipticals reaching $\mu_g\sim29$ show that these galaxies follow a S\'ersic profile with low S\'ersic index, more akin to dwarf ellipticals than giant ellipticals.
\item Spectroscopic follow-up will provide the outer halo dynamics necessary for a systematic study of the dark matter content of isolated ellpticals, which might be rather small, as shown in the case of NGC 7507 \cite{salinas12,lane15}. Given the large distances to these systems, measuring the ages and
metallicities of their GCs spectroscopically is prohibitive; therefore, these quantities will probably only be accessible through near-infrared imaging. A near-infrared imaging campaign for most of the galaxies discussed here is underway.

.
\end{itemize}

\vspace{6pt}

\acknowledgments{Based on observations obtained at the Gemini Observatory (Gemini programs: GS-2011B-Q-83, GS-2012A-Q-6), which is operated by the Association of Universities for Research in Astronomy, Inc., under a cooperative agreement with the NSF on behalf of the Gemini partnership: the National Science Foundation (United States), the National Research Council (Canada), CONICYT (Chile), Ministerio de Ciencia, Tecnolog\'{i}a e Innovaci\'{o}n Productiva (Argentina), and Minist\'{e}rio da Ci\^{e}ncia, Tecnologia e Inova\c{c}\~{a}o (Brazil).  T.R. acknowledges support from the BASAL Centro de Astrof\'{i}sica y Tecnolog\'{i}as Afines (CATA) PFB-06/2007. }

\authorcontributions{T.R and R.S. designed the observations and started this collaboration; R.R.L. reduced and analyzed the MOSAIC data; A.A. and R.S. reduced and analyzed the Gemini data, T.R. analyzed the VLT data, N.H., M.S. and R.S. reduced and analyzed the Las Cumbres data. R.S. wrote this contribution.}

\conflictsofinterest{The authors declare no conflict of interest.}

\abbreviations{The following abbreviations are used in this manuscript:\\
\noindent
\begin{tabular}{@{}ll}
GC   & globular cluster\\
GCS  & globular cluster system\\
$S_N$ & specific frequency
\end{tabular}}



\end{document}